\documentclass[aps,prl,twocolumn,superscriptaddress,groupedaddress]{revtex4}

\usepackage{amsmath}    
\usepackage{graphicx}   

\usepackage[applemac]{inputenc}

\usepackage{verbatim}   
\usepackage{color}      
\usepackage{subfigure}  

\begin{document}

\title{Critical Exponents of the Random Field Hierarchical Model}
\author{Giorgio Parisi $^{1,2}$ and Jacopo Rocchi $^1$ \bigskip\\
1. \textit{Dipartimento di Fisica, Università “La Sapienza”, P.le A. Moro 2, I-00185 Roma, Italy}
\\ 
2. \textit{INFN – Sezione di Roma1, CNR – IPCF UOS Roma}}

\begin{abstract}

We have studied the one dimensional Dyson hierarchical model in presence of a random field. This is a long range model where the interactions scale with the distance with a power law-like form $J(r) \sim r^{-\rho}$ and we can explore mean field and non-mean field behavior by changing $\rho$. Thus, it can be used to approach the phase transitions in finite-dimensional disordered models.
We studied the model at $T=0$ and we numerically computed its critical exponents in the non-mean field region for Gaussian disorder.
We then computed an analytic expression for the critical exponent $\delta$, that holds in the non-mean field region, and we noted an interesting relation between the critical exponents of the disordered model and the ones of the pure model, that seems to break down in the non-mean field region. We finally compare our results for the critical exponents with the expected ones in $D$-dimensional short range models and with the ones of the straightforward one dimensional long range model.

\end{abstract}

\maketitle

\section{Introduction}

The critical behavior of models in the presence of a quenched random field has attracted a lot of attention since the pioneering work of Imry and Ma \cite{ImryMa} because of its very interesting nature.  In fact, while in a short range system with Ising spins (RFIM) a simple domain wall argument suggests that a low temperature ordered phase can survive just if $D > 2$, the exact value of the lower critical dimension has been a debated issue for a long time.
In particular it was not clear whether or not a phase transition occurred in 3 dimensions until Imbrie demonstrated it does and $D_L^c=2$ \cite{Imbrie, Aizenman}.

This work solved a problem but left open another one: why other approaches such as perturbation theory \cite{Aharony, YoungPer} and the super-symmetric (SUSY) approach of Parisi and Sourlas \cite{ParSour} predicted the wrong result $D_L^c=3$.
It's convenient to consider the modified hyper-scaling relation that holds in presence of a random field.
In pure systems, while the entropy density of a finite volume system at infinite temperature is proportional to $\ln2$, one can argue that near $T_c$ there is a non-analytic term that is proportional to $\xi^{-D}$ because there are $V/\xi^D$ clusters that can flip, where $V$ is the finite volume and $\xi$ the correlation length. Thus, the non analytic part of the free energy density $f \sim t^{D\nu}$, where $t\sim (T-T_c)/T_c$ and $\xi \sim t^{-\nu}$, and the usual hyper-scaling law $2-\alpha=D\nu$ holds, where $\partial^{2}_{t} f  \sim t^{-\alpha}$. Anyway, in the random field model, this hyper-scaling law has to be generalized including the exponent $\theta$ \cite{Fish, Aha, Gri},
\begin{equation}
2-\alpha=\nu(D-\theta) \:,
\label{eq:modsca1} 
\end{equation}
where $\theta=2$ according to perturbation theory and the SUSY approach \cite{Aharony, YoungPer,ParSour}. 
In fact, if we denote by an overbar the average over the disorder and we call $h$ the effective random field acting on a correlated cluster at the critical point, its energy $u \sim  \overline{ m h}$ can't be neglected anymore and gives the relevant contribution to the free energy. Thus the temperature becomes an irrelevant variable as the thermal fluctuations are less divergent than the sample-to-sample ones \cite{ParSour} and the paramagnetic-ferromagnetic transition can be studied at $T=0$.

From eq. (\ref{eq:modsca1}) follows that $\theta$ acts as a dimensional reduction exponent in the sense that the phase transition in $D$ dimension has the same relations between its critical exponents of the phase transition of the pure model in $\theta$ dimension less. It's still unknown how $\theta$ decreases to one in the limit $D \rightarrow 2 ^{+}$. Anyway, a simple result can be obtained if we approximate $m$ by $\chi h$. In fact, this leads to $u \sim \chi \overline{h^2} \sim t^{-\gamma+D\nu}$ since $\overline{h^2}$ is proportional to $\xi^{-D}$ and $\chi \sim t^{-\gamma}$. Thus $\theta=\gamma/\nu=2-\eta$, where $\eta$ is defined form the connected correlation function as $C_{conn}(r)\sim r^{-(D-2+\eta)}$. This relation was proposed by Schwartz \cite{Schwartz, SchwartzSoff,SchwartzGofNatt} and other authors \cite{EisBin, VinkFishBind, VinkBind, Vink}.
Another relation may be found using a scaling theory at $T=0$ \cite{BrayMoore}, $\theta=2+\eta-\overline{\eta}$, where $\overline{\eta}$ is defined from the disconnected correlation function as $C_{disc}\sim r^{-(D-4+\overline{\eta})}$, and would imply that $\overline{\eta}=2\eta$ in order to be consistent with the previous formula.
This has been proven to be true in $D=2+\epsilon$ dimensions at the first order in $\epsilon$ \cite{BrayMoore} and numerically all simulations give a small value for $2\eta-\overline{\eta}$. The most impressive and recent one in the $3$-dimensional RFIM \cite{MayorFytas} states that $ 2\eta - \overline{\eta} \sim 10^{-3}$. 
In $D=4$, numerical studies \cite{Hartmann4D, Middleton} lead to $ 2 \eta -\overline{\eta} \sim -0.01 \pm 0.05 $ while in $D=5$, from the critical exponents computed in  \cite{Hartmann567}, it's difficult to obtain an estimation of $ 2\eta - \overline{\eta}$.
More generally, whereas or not $\theta$ is an independent exponent is not clear. A non-perturbative functional renormalization group approach \cite{TarjTiss1, TarjTiss2, TarjTiss3} suggests that the relation $\overline{\eta}=2\eta$ is not true in general.
In particular, this approach shows that for dimensions greater than $D \approx 5.1$, $\theta=2$ and $\overline{\eta}=\eta$  \cite{TarjusTissierBalog}, as can be found in \cite{Aharony, YoungPer, ParSour}. 
 
In this work we computed the critical exponents of the Dyson hierarchical version of the Random Field (RF) problem. The Hierarchical Model (HM) is a one dimensional model with a long range interaction invented by Dyson \cite{Dyson} where the interaction between spins mimics a power law decreasing potential  $J(r)=r^{-\rho}$. We will give a summary of its main features in the following section and we will refer to this model as RFHM.
In a general one dimensional long range model $\rho$ controls the distance from mean field: increasing $\rho$ the system becomes less and less mean field.
This is qualitatively similar to explore different dimensions in a $D$-dimensional short range system, where the critical behavior may be or not of the mean field type and this feature motivated many studied on disordered versions of the long range model \cite{KatzYoun, KatzLarsYoun, LarKatzMoo, LeuzziParisiRicciRuiz1, LeuzziParisiRicciRuiz2, BanosFernMartYou, ParisiLeuzzi, Hartmann}.
It must be noted that in these models the integer parameter $D$ is replaced by a continuous parameter $\rho$ and a mapping between them has been proposed in \cite{LarKatzMoo} and recently revisited in \cite{BanosFernMartYou} and \cite{ParisiLeuzzi}.
This mapping is believed to hold in the whole mean field region and near the upper critical dimension but seems to break down near the lower critical dimension \cite{Angelini Thesis}. 

We decided to study the hierarchical model instead of the long range counterpart because the two models are very similar, even if they are not thought to be in the same universality class for $\rho$ tending to the lower critical dimensions \cite{Angelini Thesis}, that is $\rho=2$ in the pure model and $\rho=3/2$ in the RF case.
The main advantage of the Dyson model respect to the straightforward long range one is that it can be studied trough an iteration equation and we don't need approximate techniques to compute equilibrium observables because the equilibrium distribution of the magnetization $P(M)$ can be exactly computed at every temperature in a polynomial time.
This iteration equation is not spoiled by the disorder induced by a random field.
The time complexity of this algorithm is $O(N^2)$, where $N$ is the size of the system, even a $T=0$.
Moreover the model can be studied at $T=0$ using a recently developed algorithm \cite{MonthusGarel} whose time complexity is $O(N \log N)$ and that computes the ground state magnetization and energy of a disordered sample. 
Thus, it gives access to big systems and accurate statistics in a reasonable computation time and this is the algorithm we used to compute critical exponents at $T=0$.

The work is organized as follows. In the first section we introduce the main features of the HM, both in its pure version and its RF version. In the second section we explain how we computed the critical exponent $\nu$ and plot the curve $1/\nu(\rho)$. In the third section we compute other critical exponents and note an interesting relation between critical exponents of the RFHM and the ones of the pure HM, somehow reminiscent of the phenomenology of the $D$-dimensional short range models. In the last section we draw the conclusions: we compare our critical exponents with the ones of the RFIM in $3$ and $4$ dimensions, using the results obtained in \cite{MayorFytas} and \cite{Hartmann4D}, and with the critical exponents of the one dimensional long range model studied in \cite{ParisiLeuzzi} and \cite{Hartmann}.

\section{The hierarchical model}
The HM is defined by \cite{Dyson}
\begin{equation}
H_{n}(s_{1},\ldots s_{N})=-\sum_{p=1}^{n}\left(\frac{c}{4}\right)^{p}\sum_{r=1}^{2^{n-p}}S_{pr}^{2}\:,
\label{eq:intro_uno}
\end{equation}
where $c$ is a coupling constant,  $N=2^{n}$ is the total number of spins and $S_{pr}$ is the sum of all the spins contained in the $r$-th $p$-level block:
\begin{equation}
S_{pr}=\sum_{i=(r-1)2^{p}+1}^{r2^{p}}s_{i}\:,\qquad\qquad r=1,\ldots,2^{n-p}\:.
\end{equation}
Spins are organized in a hierarchy of levels, indexed by $p$, whose physical meaning is that spins at the same level interact with each other through the same coupling. 
As said in the introduction, this model resembles a one dimensional chain where the interaction between spins is given by a power law decreasing potential  $J(r)=r^{-\rho}$, where $c=2^{2-\rho}$. 
It has been intensively studied since sixties \cite{BleSinai, ColletEck, KimThom, BleMaj, Jona} and its introduction by Dyson \cite{Dyson} served to demonstrate the presence of phase transitions in one dimensional problems with long range interactions. Later on, it was noted \cite {Baker, Felder} that it could be very useful in the study of the renormalization group theory developed by Wilson \cite{Wilson}.
More recently it has been studied in the field of quenched disordered models, \cite{MonthusGarel}, \cite{ParFraJor, MezParCasDec, CastPar, Cast, RicciAngPar, RodBray}, as well as for the Anderson localization problem \cite{MonthusGarelAndLoc,MetzPar}.

\subsection{Pure model}

The interesting interval in which $\rho$ can take values is between one and two.
In fact, for $\rho<1$ the free energy corresponding to eq. (\ref{eq:intro_uno}) is not defined in the thermodynamic limit; the limit $\rho \rightarrow 1^{+}$ corresponds to the limit $D\rightarrow \infty$ in short range theories.
On the other hand, for $\rho>2$ there is not a phase transition \cite{Dyson}. One of the way to verify this statement is to see that the singular part of the cost of a bubble in a magnetized phase is of order $L^{2-\rho}$ and so  bubbles have $O(1)$ cost for $\rho>2$. The non trivial critical region, where critical exponents differs from their mean field values, is $\rho \in (3/2,2)$ \cite{BleMaj, KimThom, ColletEck}. 
This may be seen from the the hierarchical structure of the Hamiltonian in eq. ($\ref{eq:intro_uno}$), that allows an exact realization of the block spin transformation \cite{Kadanoff, Kadanoffetal}.

Let's write the partition function,
\begin{equation}
\mathcal{Z}_{N} =\int  d s_{1} \ldots ds_{2^{n}} \exp \left \{-\beta H_{n}(s_{1},\ldots s_{N})  + \sum_{i=1}^{2^n}f(s_{i})\right\}\:, \nonumber
\end{equation}
where $P(s)=\exp\{f(s)\}$ is a weight function on each spin $s$,  for example the Ising weight $\delta(s^2-1)$. 
After the RG transformation
\begin{equation}
\frac{s_{i}+s_{i+1}}{2}=\gamma s_{(i+1)/2}'\quad\frac{s_{i}-s_{i+1}}{2}=t_{(i+1)/2}'\:,
\label{eq:kad_lu}
\end{equation}
$\mathcal{Z}_{N}$ may be rewritten in terms of the new effective spins $\left \{ s'_{i} \right \}_{i=1,\ldots, 2^{n-1}}$  through the integration over the other $2^{n-1}$ variables $\left \{ t '_i \right \}_{i=1,\ldots, 2^{n-1}}$. 
The number of degrees of freedom has been halved, and the prize that has been paid is the introduction of a new weight function $P'(s')$, given in terms of the old one by 
\begin{equation}
P'(s')=e^{4\beta J\gamma^{2}s'^{2}}\int d t ' P\:(\gamma s'+t ')P\:(\gamma s'-t ')\:,
\label{eq:intro_4}
\end{equation}
where $J=c/4=2^{-\rho}$.
It's worth to notice that this equation has the same form of the approximate recursion formula derived by Wilson \cite{Wilson,Baker,Felder,Polch}. In this sense, HM is a model for which Wilson's formula is exact.
A part form the new weight function in eq. (\ref{eq:intro_4}) the new Hamiltonian has the hierarchical structure of the old one and a new coupling constant: if $K=\beta J$, $K'$ is given by $K'=4J\gamma^{2}K$ and it depends on $\gamma$.
At the critical point, where the interaction between clusters of spins does not change with the scale at which we observe the system, we impose
\begin{equation} 
\gamma=2^{\rho/2-1}\:
\label{eq:intro_5}
\end{equation}
and it's natural to suppose that eq. (\ref{eq:intro_4}) has a fixed point. We can use this result to an evaluate the $\eta$ index, defined as $C_{conn}(r)=\left< s_i s_{i+r} \right> -\left< s_i\right> \left<s_{i+r} \right>\sim r^{-(D-2+\eta)}$. In fact, suppose that the above transformation is iterated $n$ times:
\begin{equation}
\frac{\sum_{i}^{2^n}s_{i}}{2^{n}}=\gamma^{n}s_{i}^{(n)} \nonumber
\end{equation}
where $s_{i}^{(n)}$ is the renormalized spin after $n$ transformations. $\gamma$ absorbs the diverging part of the r.h.s. of the last equation, making  $s_{i}^{(n)}$ a finite quantity. Thus we obtain
\begin{equation}
m= \frac{\left< \sum_{i}^{2^n}s_{i} \right> }{2^{n}}   \propto N^{\rho/2-1} \:, 
\end{equation}
and since $C_{conn}(r) \sim m^2 $ near $T_c$, the susceptibility scales as
\begin{equation}
\chi_{conn}=\sum_{r} C_{conn} (r) \propto N\gamma{}^{2n} \propto N^{\rho-1}\:.
\label{eq:intro_dueee}
\end{equation}
In a $1$-dimensional finite size system it may also be expressed as 
\begin{equation}
\chi_{conn,L}(T=T_{c})=\int_{L}\frac{dr}{r^{D-2+\eta}} \propto L^{2-\eta}\qquad D=1,
\label{eq:intro_treeee}
\end{equation}
where $L=N$ is the size of the system, and thus, comparing eq. (\ref{eq:intro_dueee}) and eq. (\ref{eq:intro_treeee}) we obtain $\eta(\rho)=3-\rho\:$. This relation holds both in mean field and non mean field regions, the reason being that we performed an exact Kadanoff  transformation and computed an exact value for $\gamma$ at $T_c$.

From eq. (\ref{eq:intro_4}) and eq. (\ref{eq:intro_5}) we can also calculate $\nu$ but we must do an ansatz on the form of $P(s)$, and then we must study its stability. If $P(s)$ is a normal distribution $\mathcal{N}(0,1)$, the new weight function is still a normal distribution whose variance $(\Sigma')^2$ is given by
\begin{equation}
\frac{1}{2(\Sigma')^{2}}=\frac{1}{c\Sigma^{2}}-\beta\:, \qquad \Sigma=1\:.  
\label{eq:intro_55}
\end{equation}
The only unstable fixed point of the recursion equation (\ref{eq:intro_55}) is found imposing $\Sigma'=1$, i.e. when the system is invariant under RG transformations. This lead to an evaluation of the critical temperature $\beta_{c}=\frac{2-c}{2c}$ for this particular choice of the weight function and to the fixed point value $O^*=1/2$  for the operator $O=1/(2\Sigma)$. $\nu$ can be extracted from the evolution of a small perturbation from this fixed point value, that is starting with $O=1/2+\delta$ and calculating $\delta'$. From eq. (\ref{eq:intro_55}) we have
\begin{equation}
\frac{1}{2}+\delta'=2^{\rho-1}\left(\frac{1}{2}+\delta\right)-\frac{2-c}{2c}\:  \nonumber
\end{equation}
and thus $\delta'=2^{\rho-1}\delta$, leading to $\nu^{-1}(\rho)=\rho-1$. While the previous calculated expression for $\eta$ holds whatever $\rho$, this expression for $\nu$ is valid when the Gaussian ansatz is stable. Many perturbative analysis has been done \cite{BleMaj,KimThom, ColletEck} and it has been found that the mean field region is $\rho \in (1,3/2)$.
An easy way to grab the upper critical value of $\rho$ is to use hyper-scaling relations, for example $2-\alpha=\nu D$, with $D=1$, since they are valid just in the non-mean field region \cite{ParisiBook,Coniglio} up till the upper critical dimension, where they are satisfied by classical indices:
\begin{equation}
2=\frac{1}{\rho^u_c-1} \Longrightarrow \rho^u_c=\frac{3}{2}\:. 
\end{equation}

\subsection{Random field model}

We now consider the case in which there are uncorrelated random fields whose variance is $h^2$. The effect of the random fields is to weaken the ordered phase and thus it's natural to expect that it survives just in regions where $T$ and $h$ are small. A simple domain wall argument implies that the singular part of the cost of a bubble is of order $L^{2-\rho}-h^2 L^{1/2}$: a low temperature-low disorder magnetized phase may survive as long as $\rho \in (1,3/2)$. The non trivial region is instead given by $\rho \in (4/3, 3/2)$ as found by Rodgers and Bray \cite{RodBray}. 

In order to deal with the disorder we replicate the partition function
\begin{equation}
\begin{split}
\mathcal{Z}_{N}^{m}&=\int\prod_{\alpha}\prod_{i=1}^{2^{n}}ds^{\alpha}_{i}\exp\left\{ \sum_{i\alpha}f(s_{i}^{\alpha})+ \right. \\
& \left. -\beta H_{n}(s^{\alpha}_{1},\ldots s^{\alpha}_{N}) +\beta\sum_{i}h_{i}\sum_{\alpha}s_{i}^{\alpha}\right \}
\end{split}
 \nonumber
\end{equation}
where $\alpha$ runs over the $m$ replicas and $\overline{h_i h_j}=h^2 \delta_{ij}$. The next step is to average over the disorder, assuming it is Gaussian,
\begin{equation}
\begin{split}
\overline{\mathcal{Z}_{N}^{m}}&=\int\prod_{\alpha}\prod_{i=1}^{2^{n}}ds^{\alpha}_{i} \:\exp\left \{ \sum_{i\alpha}f(s_{i}^{\alpha}) + \right. \\ 
& \left. -\beta H_{n}(s^{\alpha}_{1},\ldots s^{\alpha}_{N}) + \frac{h^2 \beta^{2}}{2} \sum_{i,\alpha\beta}s_{i}^{\alpha}s_{i}^{\beta}   \right \} \:
\end{split}
\nonumber
\end{equation}
and make the same RG transformation as before, eq. (\ref{eq:kad_lu}). Again, the partition function may be written in terms of the new effective spins $\left \{ s'_{i} \right \}_{i=1,\ldots, 2^{n-1}}$, if we introduce a new weight function
\begin{equation} 
\begin{split} 
& P'\left( \{ s'_{\alpha} \} \right) = e^{4\beta J\gamma^{2} \sum_{\alpha}(s'_{\alpha})^{2}}  \int\prod_{\alpha} d t '_{\alpha} \prod_{\alpha}   \\ 
& P(\gamma s'_{\alpha}+ t '_{\alpha}) P(\gamma s'_{\alpha}-\ t '_{\alpha})  e^{\beta^{2} h^2 \sum_{\alpha\beta} t '_{\alpha}t '_{\beta} }\:.
\end{split}
 \label{eq:newrfrec}
\end{equation}
The new coupling constant is $K'=4J\gamma^{2}K$ and there is also a new variance $(h')^2=2 \gamma^2 h^2$.
At the ferromagnetic critical point $T/J$ is invariant under RG transformations, leading to eq. (\ref{eq:intro_5}), and this gives $(h ')^2=2^{\rho-1} h^2$. It means that this fixed point is always unstable respect to the addition of a random field. The local renormalization group flow departs from there versus regions of higher disorder and thus at the relevant RF-fixed point one expects that $h/J$ is invariant, even if nor $J$ nor $h$ are. This may only happen if $T=0$. Moreover, the RG invariance of $h/J$ implies that the value $\gamma$ at this fixed point is given by
\begin{equation}
\gamma=2^{\rho-3/2}\:.
\label{eq:gamma_rffff4a}
\end{equation}
This relation leads to
\begin{equation}
m^{(2)}=\frac{1}{{2^{2n}}} \overline{ \left( \left< \sum_{i}^{2^n}s_{i} \right> \right)^2 }  \propto N^{2\rho-3}\:,
\label{eq:def_m2222q}
\end{equation}
that can be used to evaluate $\eta$ and $\bar{\eta}$. This last critical index is defined from the disconnected correlation function $C_{disc}(r)=\overline{\left<s_i\right> \left<s_{i+r}\right>}$ as  $ C_{disc}(r)\sim r^{-(D-4+\bar{\eta})}$ for $r \gg1$. Thus 
\begin{equation}
\chi_{disc}=\sum_{r}C_{disc}(r) \propto N m^2 \propto N^{2\rho -2}\:.
\label{eq:rf_cdisc1}
\end{equation}
In a 1-dimensional system it can also be expressed as
\begin{equation}
\chi_{disc,L}(h=h_c)=\int_L \frac{dr}{r^{D-4+\bar{\eta} } } \sim L^{4-\bar{\eta}}
\label{eq:rf_cdisc2}
\end{equation}
and thus comparing eq. (\ref{eq:rf_cdisc1}) and eq. (\ref{eq:rf_cdisc2}) we get $\overline{\eta}(\rho)=6-2\rho$.  In RF models $\eta$ is still defined from the connected correlation function as $C_{conn}(r)=\overline{\left< s_i s_{i+r} \right> -\left< s_i\right> \left<s_{i+r} \right>}$ and thus at the critical point $\chi_{conn,L}$ scales  with $N$ as in eq. (\ref{eq:intro_treeee}). This equation may be now compared to the other definition of the connected susceptibility $\chi_{conn}=T d m / d H $ , where $H$ is a uniform field that gets renormalized according to $H'=2 \gamma H$ and $T\sim J$ in the region $T/J \ll 1$.
Equation (\ref{eq:gamma_rffff4a}) implies that $J'=2^{\rho-1} J$ at the zero temperature fixed point and after $n$ steps of the renormalization procedure, we get $H\sim (2\gamma)^{-n}\sim N^{1/2-\rho}$, while the coupling constant scales as $J \sim N^{1-\rho}$.
This leads to
\begin{equation}
\chi_{conn}=T\frac{d m}{d H}\sim J \frac{d m}{d H} \sim N^{\rho-1}\:.
\end{equation}
From eq. (\ref{eq:intro_treeee}) we get $\eta(\rho)=3-\rho$, in the same way as in the pure model. Therefore $\eta=\bar{\eta}/2$, \cite{RodBray}, leading to $\theta(\rho)=\rho-1$ for every $\rho$. 
The same result for $\theta$ can be obtained looking at the RG exponent associated to $J$. In fact, the scaling theory at $T = 0$ developed by Bray and Moore \cite{BrayMoore}, assumes that $J$ gets renormalized even at the critical point, but that the ratio $h/J$ is invariant, and $\theta$ enters in this calculation as the RG exponent associated to $J$: $J'=b^{\theta}J$, where $b$ is the rescaling factor. 

If we do a Gaussian ansatz for the weight function, eq. (\ref{eq:newrfrec}) implies that the mean field value of $\nu$ is given by $1/(\rho-1)$ \cite{MonthusGarel}, \cite{RodBray}, as in the pure case. The non-mean field region is given by $\rho \in (4/3, 3/2)$ and again hyper-scaling relations, e.g. eq. (\ref{eq:modsca1}), may be used to compute the upper critical value of $\rho$ since they are valid only in non-mean field regions and are satisfied at the upper critical dimension by classical indices:
\begin{equation}
2=\frac{2-\rho^u_c}{\rho^u_c-1} \Longrightarrow \rho^u_c=\frac{4}{3}\:.
\end{equation}

\section{Calculation of $\nu$}

The hierarchical structure of the Hamiltonian implies that 
\begin{equation}
\begin{split}
H_n(s_1,\ldots,s_{2^n})=&H^{(L)}_{n-1}(s_1,\ldots, s_{2^{n-1}})+ \\ 
&H^{(R)}_{n-1}(s_{2^{n-1}+1},\ldots, s_{2^n})+\Delta_{L,R}^{n}\:,
\end{split}
\end{equation}
where $N=2^n$ is the number of spins,
\begin{equation}
\Delta_{L,R}^{(n)}=-J_{n}\left(\sum_{i=1}^{2^n}s_i\right)^{2}
\end{equation}
is the interaction term and $J_{n}=(c/4)^n=(1/2^{\rho})^n$. 
This relation is not spoiled by a random field, the only difference respect to the pure case is in the starting condition. In the pure case, $P_0(s)=1/2 [\delta(s-1)+\delta(s+1)]$ while in presence of disorder $P_0(s)=\exp[\beta h s ] / 2\cosh (\beta h)$ where $h$ is the random field acting on $s$.
This equation may be used to find a recursion equation for the probability $P_l(M)$ that a system of $2^l$ sites has magnetization given by $M$. $P_l(M)$ is defined by
\begin{equation}
P_l(M)\sim \sum_{\{s_i\}_{i=1,\ldots, 2^l}} \delta \left(M -\sum^{2^l}_i s_i \right) e^{-\beta H_l (s_1,\ldots,s_{2^l}) }\:,
\end{equation}
where the symbol $\sim$ means that a normalization factor is understood, and the recursion equation obeyed by the probabilities reads
\begin{equation}
P_l(M) \sim  e^{\beta J_l M^2} \sum_{   S=-2^{l-1}   }^{     2^{l-1}    } P_{l-1}(S) P_{l-1}(M-S)\:.
\label{eq:receqqqq}
\end{equation}
The sum is done over all the possible magnetizations of the smaller systems, that is $S=\{-2^{l-1},-2^{l-1}+2,-2^{l-1}+4, \ldots, 2^{l-1} \}$. 
This equation may be used to calculate the moments of the distribution $P_l(M)$ up till the last level of the interaction, it holds whatever $T$ and $h$ and thus it is suitable to study the transition at $T>0$ both fixing the temperature and tuning the variance of the random field and viceversa.
It is widespread believed that the critical behavior would be the same and would be the same of the transition at $T=0$.

As we said before, we decided to study the $T=0$ transition using the algorithm that Monthus and Garel have recently developed \cite{MonthusGarel}.
This algorithm finds the ground state of a RFHM sample in a linear time, a part from logarithmic corrections, and it is faster than the $T=0$ limit of eq. (\ref{eq:receqqqq}), whose time complexity is quadratic.
It is based on the observation that in presence of an external uniform field $H$, the energy of a configuration $C$ is linear in $H$: $E(C)=-M_C H + a_C$ and that as $H$ grows, the ground state magnetization also grows. This is called "no-passing rule" \cite{FronVives,LiuDam} and in other words, the collection of ground states for different values of $H$ is ordered in magnetization: the magnetization of the ground state is an nondecreasing function of $H$.
The magnetic field at which there's a change in the ground state corresponds to a collective rearrangement of spins. This phenomenon is called equilibrium avalanche.
Given a configuration made by $2^l$ sites, we can divide it in its left and right part, each one containing $2^{l-1}$ sites. The relation between $a_{C_L}$, $a_{C_R}$ and $a_C$ is \cite{MonthusGarel}:
\begin{equation}
a_C=a_{C_L}+a_{C_R}-J_l M_C^2\:, \qquad M_C=M_L+M_R,
\end{equation}
and thus, given the ground state list of the left and of the right part, we can construct the ground state list of the total system versus $H$. 
These lists contain no more than $2^l$ configurations \cite{MonthusGarel} and thus this procedure takes a $O(N)$ time. We have checked for several samples that the ground state magnetizations computed using the $T=0$ limit of eq. (\ref{eq:receqqqq}) and with this faster algorithm are the same and since we have to average over many disordered samples, this is the best option to study the $T=0$ transition.

\begin{figure}
\includegraphics[scale=0.3]{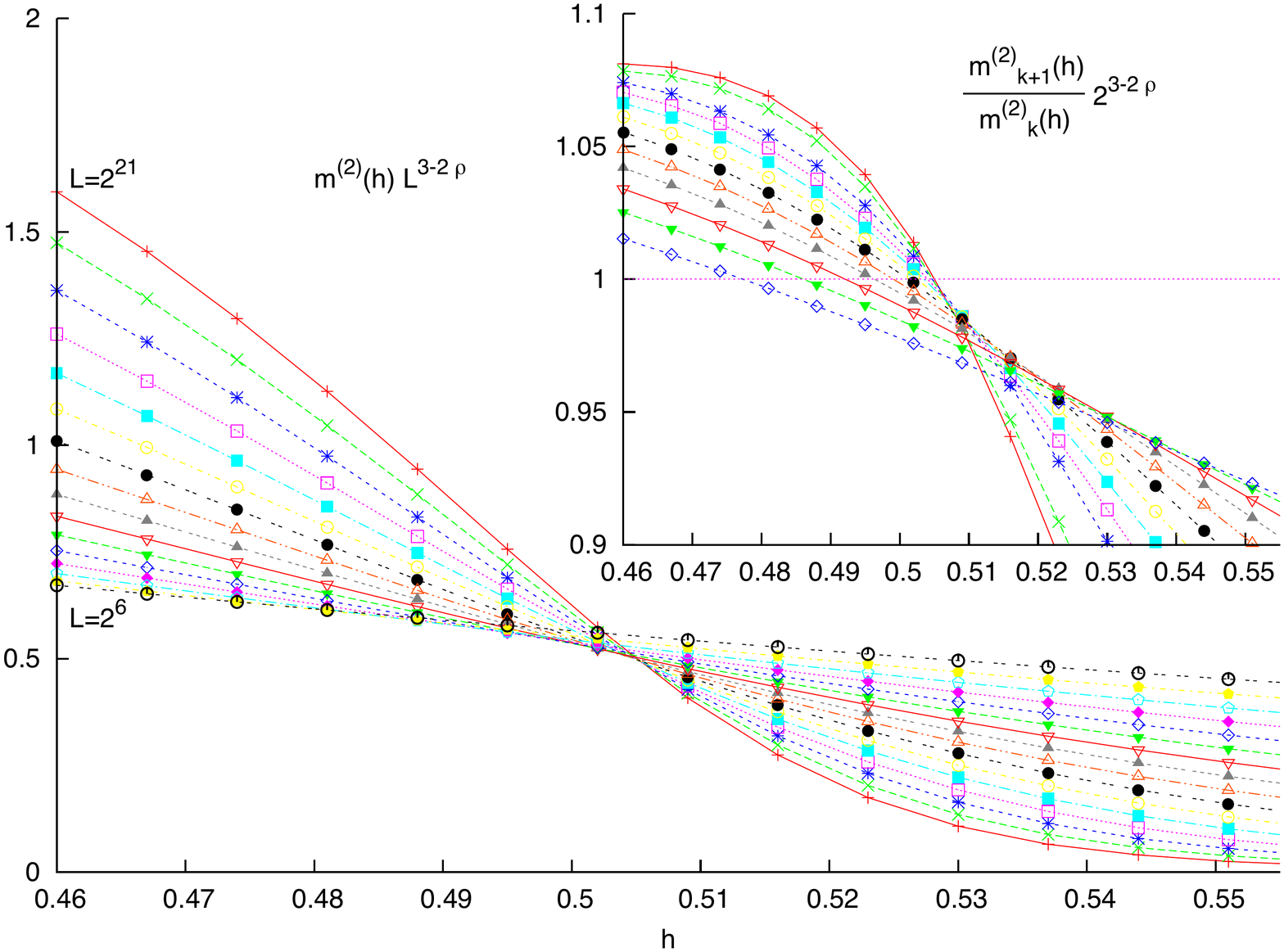}
\caption{\label{fig:m2} Plot of the curves $O_k(h)$ versus $h$, where $O_k(h)$ is defined in eq. (\ref{eq:OOO}). $h_c$ is given by the point where different curves crosses. Inset: we plotted $O_{k+1}(h)/O_k(h)$ in order to show that as $k$ grows, this ratio goes to one as it should, being $O_k(h)$ a size invariant quantity. In this plot $\rho=1.465$. Lines are a guide for the eye.}
\end{figure}

\begin{figure}
\includegraphics[scale=0.7]{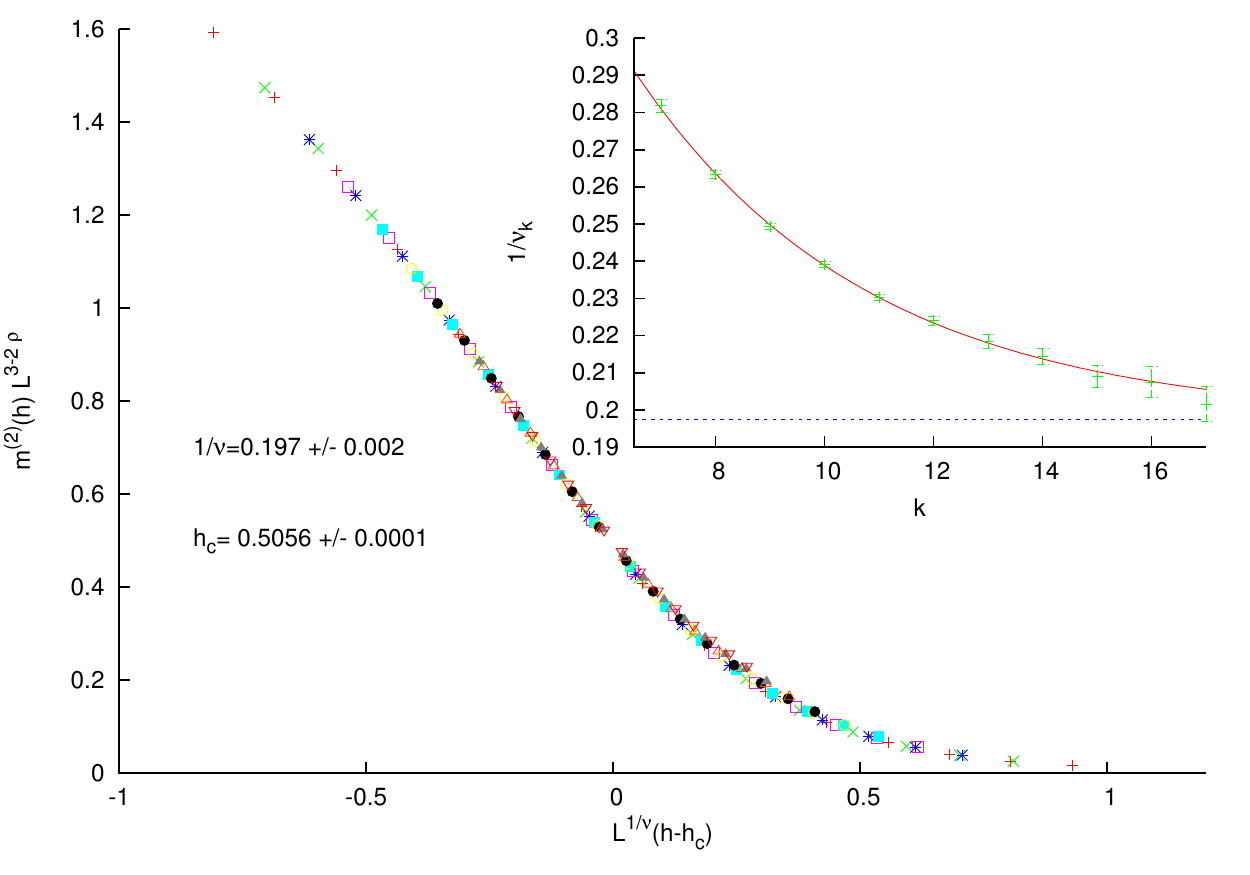}
\caption{\label{fig:nu} Collapse of the curves $O_k(h)$ versus $L^{1/\nu}(h-h_c)$ for $\rho=1.465$. Good collapse occurs just for big enough systems, and thus here we have taken $k$ from $13$ to $21$.  
Inset: plot of the curve $\nu_k$ versus $k$, computed as in eq. (\ref{eq:nukdef}), to get the asymptotic value of $\nu$. The errors over $\nu_k$'s have been computed with the \textit{bootstrap} method as explained in the text. We found 
$1/\nu=0.197(2)$ and $\omega=0.33(3)$, see eq. (\ref{eq:fss_nu}).
The red line serves just as a guide for the eye.
In this plot $\rho=1.465$.}
\end{figure}

For each $\rho$ we studied around $\mathcal{N}=80000$ samples at different values of $h$ (typically we studied $15$ or more values of $h$). For each sample of size $N=2^n=2^{21}$ we extracted $\mathcal{N}_k=2^{n-k}$ ground state magnetizations of systems whose size is $N_k=2^k$.
We have considered $k \in (6, 21)$. The hierarchical structure of the Hamiltonian makes it possible to divide a sample in two subsamples and calculate their magnetizations before considering the coupling between spins of the two different parts. Each subsample can be further divided and this procedure can be repeated until the single spins. So, at the end, we have much more samples to average over for small sizes than we have for bigger sizes.
For each $h$ and $k$ we have then randomly picked up $n_k/2$ samples and averaged their squared ground state magnetization, repeating this procedure $M$ times. 
We have found that $M\sim50$ was already big enough. Each time, we have used these data to compute the observable $O_k(h)=L^{2y}m^{(2)}_k(h)$, where $y=3/2-\rho$ and $m^{(2)}$ has been defined it eq. (\ref{eq:def_m2222q}). This quantity is size invariant at the critical point, see eq. (\ref{eq:gamma_rffff4a}), and thus the intersections of these curves, for different values of $k$, gives the critical value of $h$ \footnote{
Respect to the model studied in \cite{MonthusGarel} the couplings $J_k$ have been rescaled with the factor $1-2/2^{\rho}$, thus their critical variance is equal to ours times this factor. This has been done in order to reduce the spread of the region in $h$  where crosses between curves occur. 
}, see Fig. \ref{fig:m2}.
The derivative $\partial O_{k+1}/\partial O_k$ at the critical point leads to $\nu_k$, the values of $\nu$ at the level $k$, in a way that will be exposed later, see eq. (\ref{eq:nukdef}) and eq. (\ref{eq:fss_nu}). The errors over $\nu_k$ has been calculated from the  standard deviation of the $M$ instances of these quantities, and the asymptotic behavior of the $\nu_k$ has been studied to get $\nu=\lim_{k\rightarrow \infty} \nu_k$, see Fig. \ref{fig:nu}.

The asymptotic values of $\nu$ may be computed from finite size scaling \cite{Cardy,MartinMayorAmit,Meurice,ParisiBook} as follows. In a system of linear size $L$, a size invariant quantity at the critical point has the form
\begin{equation}
\begin{split}
&O(L,t)=f(t L^{1/\nu}, L^{-\omega})=O_c+\\
&+f'_1 t L^{1/\nu}+f'_2 L^{-\omega}+ f''_{12} t L^{1/\nu-\omega} + \ldots\:,
\end{split}
\label{eq:ffs1}
\end{equation}
where $t$ is the rescaled difference from the critical temperature $t=(\beta-\beta_c)/\beta_c$, $L$ is the size of the system, $\omega$ is the correction-to-scaling exponent, $\nu$ is defined from $\xi \sim t^{-\nu}$ and $O_c$ is the critical value of $O$.
In the $T=0$ transition of the random field model, the only relevant variable is the rescaled difference from the critical variance, so eq. (\ref{eq:ffs1}) is still valid if we replace $t$ with  $t=(h-h_c)/h_c$. 
In eq. (\ref{eq:ffs1}) non linear terms in $t$ can be neglected since we assumed to be near the critical point. 
A scale-invariant quantity has the property to remain constant under Renormalization Group transformations at the critical point and thus, in different size systems, $O_c$ is a universal value that does not scale with $L$.
The value of $h$ at which $O(L,t)$ and $O(L',t)$ crosses is defined by $h_L^{*}$ and goes to $h_c$ as $L$ grows. Thus we have 
\begin{equation}
\frac{\partial O_{L'}}{\partial O_L}=\frac{\partial O_{L'}}{\partial h} \frac{\partial h}{\partial O_L}=\frac{f'_1L'^{1/\nu}+f''_{12}L'^{1/\nu-\omega} }{f'_1L^{1/\nu}+f''_{12}L^{1/\nu-\omega}}\:. \nonumber
\end{equation}
and taking logs on both sides we obtain
\begin{equation}
\left. \log_b \left( \frac{\partial O_{L'}}{\partial O_L} \right) \right|_{h^{*}_L}=\frac{1}{\nu} + A L^{-\omega}\:,
\label{eq:fss_nu}
\end{equation}
where $b$ is defines as $L'=b L$ and $A$ a constant.
The l. h. s. of this equation gives $\nu_k$:
\begin{equation}
\nu_k= \left. \log_b \left( \frac{\partial O_{k+1}}{\partial O_k} \right) \right|_{h_k^{*} }\:.
\label{eq:nukdef}
\end{equation}
where we recall that $k=\log_2(N_k)$. We have used these equations to compute $\nu$, with the scale invariant quantity 
$O(k,t)$ defined by
\begin{equation}
O(k,t)=L^{2y}m^{(2)}_k (h)\:,\qquad  y=\frac{3}{2}-\rho\:, 
\label{eq:OOO}
\end{equation}
where $k=6, \dots, 17$ and $b=2$. We have studied even bigger samples, until $k=21$, but the error bars for $k>17$ are usually too big to be significative.
In each of the $M$ extractions of data, we have computed the asymptotic value of the quantity in eq. (\ref{eq:nukdef}), let's call it $\nu^S$ where $S=1,\ldots,M$, and we computed $\nu$ and its error as the mean and standard deviation of the histogram of the $\nu^S$. The procedure here illustrated is called \textit{bootstrap} \cite{MartinMayorAmit}, \cite{ParisiBook}. 

\begin{figure}
\includegraphics[scale=0.7]{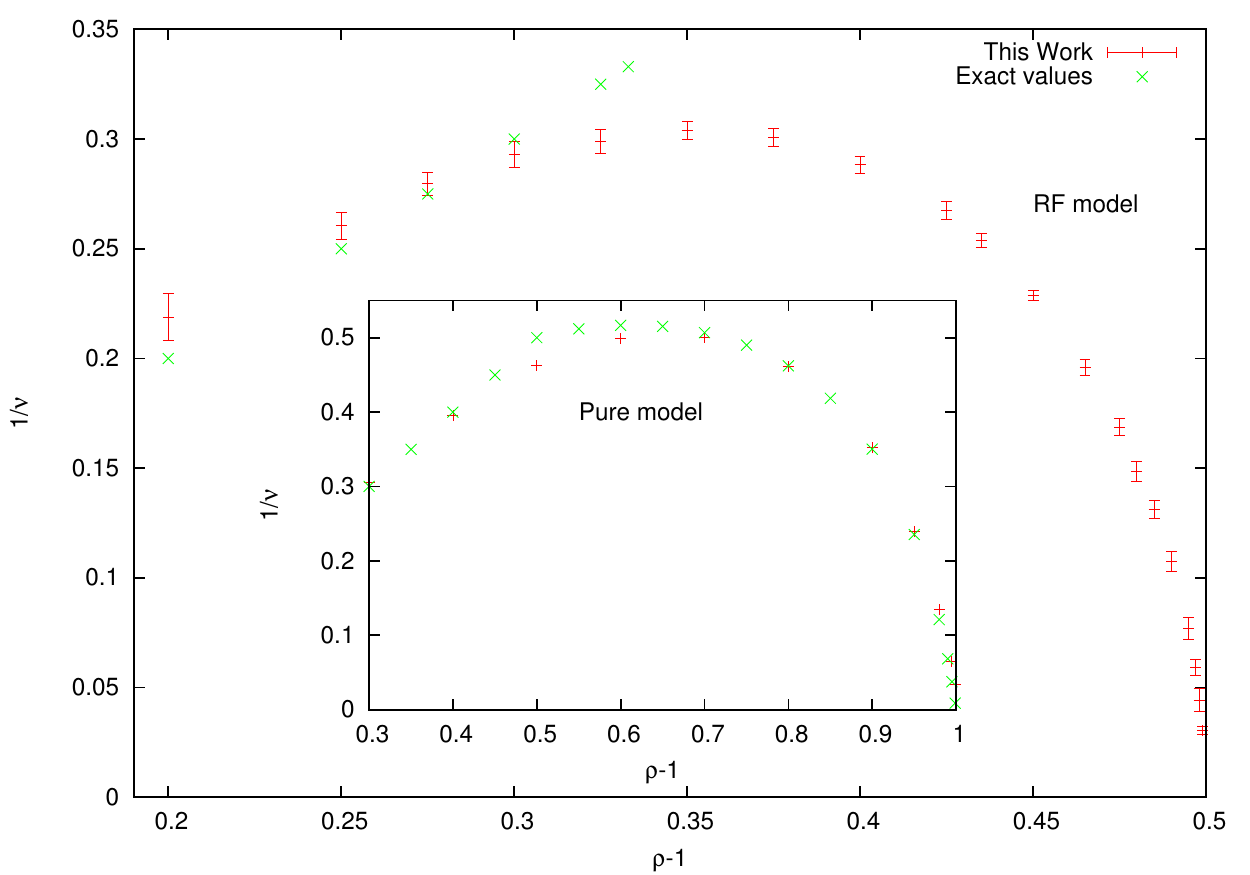}
\caption{\label{fig:nu_RF_and_Pure} 
Inverse of the critical exponent $\nu$ as a function of $\rho$ in the RFHM. The red points stands for the values of $1/\nu$ computed in this work, while the green ones are the known mean field values. 
In the disordered case the non-mean field region ($\rho>4/3$) was unexplored and here $1/\nu$ has been computed using the algorithm developed in \cite{MonthusGarel}. 
Inset: inverse of the critical exponent $\nu$ as a function of $\rho$ in the pure HM. Green points have been computed by Kim and Thompson who also computed the critical indices of the pure model in the non-mean field region ($\rho>3/2$) \cite{KimThom}. Here we confront these values with the ones we computed using eq. (\ref{eq:receqqqq}) and eq. (\ref{eq:nukdef}) to show that they work well almost in the whole non-mean field region, a part from the extremes.
Note that the relevant regions of these pictures are the non-mean field ones, as in the mean field regions $1/\nu$ is known to be equal to $\rho-1$ in both cases.}
\end{figure}

In Fig. (\ref{fig:nu_RF_and_Pure}) we show the values of $1/\nu$ computed at various $\rho'$s. We have also plotted the results we get for the pure model, that can be compared with the ones obtained in \cite{KimThom}.
The pure model can be studied using eq. (\ref{eq:receqqqq}) and eq. (\ref{eq:nukdef}).
In the pure case, as well as for the disordered case, we studied systems up to $2^{17}$ spins and the asymptotic critical exponents have been computed using eq. (\ref{eq:nukdef}), see also the inset of Fig. \ref{fig:nu}. The pure model has been studied using eq. (\ref{eq:receqqqq}) and we see that eq. (\ref{eq:nukdef}) works quite well in the non-mean field region, apart form the limits $\rho \rightarrow 3/2$ and $\rho\rightarrow 2$, where we have seen that logarithmic corrections have to be taken into account.
Thus, apart from the regions where $\rho \rightarrow 4/3$ and $\rho \rightarrow 3/2$ we expect that the values we have obtained are the correct ones.

\section{Relations between critical exponents}
In this section we obtain all the other critical exponents of the RFHM. $\delta$, defined from the vanishing of the magnetization in presence of a magnetic field $H$, $m \sim H^{1/\delta}$, has an analytical expression, shown in eq. (\ref{eq:tre_delta}). All the other critical exponents depends only on $\nu$. 
We first review the critical exponents of the pure HM.

The hyper-scaling relation $\alpha=2-D\nu$ can be used to compute the non-mean field value of $\alpha$, where $\alpha$ is defined as $\partial^2_t f \sim t^{-\alpha}$. Since $D=1$, we have
\begin{equation}
\alpha(\rho)=2-\nu(\rho)\:.
\end{equation}
While this relation gives only the non-mean field value of $\alpha$, that is zero otherwise, the other scaling relation $\gamma=\nu(2-\eta)$ is more general and it's also valid in the mean field region. $\gamma$ is defined form the divergence of susceptibility at the critical temperature, $\chi \sim t^{-\gamma}$ and it is given by
\begin{equation}
\gamma(\rho)=\nu(\rho)(\rho-1)\:
\label{eq:uno_fit}
\end{equation}
whatever $\rho$ and leads to $\gamma=1$ in the mean field region.
Moreover it's a general fact that $\delta$ only descends from $\eta$ via the relation $\delta=(D+2-\eta)/(D-2+\eta)$
and thus we have an analityc expression for  $\delta$ in the non-mean field region:
\begin{equation}
\delta(\rho)=\frac{\rho}{2-\rho}\:.
\label{eq:fit_delta_pure}
\end{equation}
From the relation $\gamma=\beta(\delta-1)$ we can also estimate $\beta$, defined from $m\sim t^{\beta}$, that turns out to be equal to 
\begin{equation}
\beta(\rho)=\left(1-\frac{\rho}{2}\right)\nu(\rho)\:.
\end{equation}
These critical exponents have been computed by Kim and Thompson \cite{KimThom} who tabulated very accurate estimations of the values of $\nu(\rho)$ (see the inset of Fig. \ref{fig:nu_RF_and_Pure}) in non-mean field region and also gave the analytic expression for $\delta(\rho)$ in eq. (\ref{eq:fit_delta_pure}).

In the RFHM the hyper-scaling law is modified according to $\alpha=2-(D- \theta)  \nu$, where $\theta$ has been defined in eq. (\ref{eq:modsca1}) and we have $\theta(\rho)=2+\eta(\rho)-\overline{\eta}(\rho)=\rho-1$. Thus, while the non-mean field critical exponent $\alpha$ is given by
\begin{equation}
\alpha(\rho)=2-(2-\rho) \nu(\rho)\:,
\label{eq:due_fit}
\end{equation}
equation (\ref{eq:uno_fit}) governing the behavior of  $\gamma$ is still valid. 
In presence of a random field $\delta$ can be expressed in terms of $\eta$ and $\theta$ according to the relation $\delta=(D-2\eta+\overline{\eta})/(D-4+\overline{\eta})$ and thus we get
\begin{equation}
\delta(\rho)=\frac{1}{3-2\rho}\:
\label{eq:tre_delta}
\end{equation}
in the non-mean field region.
At last, using again the scaling relation $\gamma=\beta(\delta-1)$ we obtain 
\begin{equation}
\beta(\rho)=\left(\frac{3}{2}-\rho \right) \nu (\rho)\:.
\end{equation}
This equation is consistent with eq. (\ref{eq:def_m2222q}) because in a finite size system $m^{(2)} (L) \sim L^{-2\beta/\nu}$ at the critical point (see also eq. (\ref{eq:OOO}) and the inset of Fig. \ref{fig:nu}).
Let us remark that we have an exact result for the critical index $\delta$, as well $\eta$ and $\overline{\eta}$, descends from the rescaling factor defined in eq. (\ref{eq:kad_lu}). 

It must be noted that eq. (\ref{eq:tre_delta}) reduces to eq. (\ref{eq:fit_delta_pure}) if we replace its argument $\rho$ by $2-1/\rho$, that is
\begin{equation}
\delta_{RF} (2-1/\rho)=\delta_{Pure} (\rho)
\label{eq:mapping}
\end{equation}
So we could ask if the rule 
\begin{equation}
\rho_{RF} \rightarrow \frac{1}{2-\rho_{RF}}\:,
\label{eq:mappingrrf}
\end{equation}
also holds for the other critical exponents. 
From our estimations of $1/\nu$ in the RFHM and from the ones made by Kim and Thompson \cite{KimThom} for the pure HM, we can compare the respective exponents $\gamma(\rho)'s$ obtained through eq. (\ref{eq:uno_fit}). 
So if eq. (\ref{eq:mappingrrf}) holds also for $\gamma$, we should have that $\gamma_{Pure} (3/2)=\gamma_{RF} (4/3)$. In fact, this is true because these two values correspond to their respective mean field thresholds and $\gamma_{Pure} (3/2) = \gamma_{RF} (4/3)=1$, as can be seen from eq. (\ref{eq:uno_fit}). 
What happens in the first parts of the non-mean field regions is not clear from our data, because as we already said they are not so good at the extremes of this region (see Fig. \ref{fig:nu_RF_and_Pure}). We have then used the perturbative results obtained in \cite{KimThom} and \cite{RodBray}, that give
\begin{equation}
\begin{split}
&\nu_{Pure}(\rho=\frac{3}{2}+\epsilon)=2-\frac{4}{3}\epsilon \\
&\nu_{RF}(\rho=\frac{4}{3}+\epsilon)=3
\end{split}
\end{equation}
at the first order in $\epsilon$. This expansions may be used to compare $\gamma_{RF}(2-1/\rho)$ and $\gamma_{Pure}(\rho)$ in $\rho=3/2+\epsilon$. It turns out that they are both equal to $1+(4/3) \epsilon$ and thus that eq. (\ref{eq:mappingrrf}) also holds for $\gamma$'s in perturbation theory, at least at the first order in $\epsilon$.

When we depart from the mean field threshold values, our data seem to suggest that this formula breaks down. For example, at $\rho_{RF}=1.45$, that would correspond to $\rho_{Pure}\approx 1.818$, a spline interpolation of the data obtained in \cite{KimThom} gives $\gamma_{Pure} (1.818) = 0.548(1)$ while according to eq. (\ref{eq:mappingrrf}) we should get $\gamma_{RF}(1.45) = 0.507(3)$.
Similar discrepancies are found also for other points in the middle of the non-mean field region, where our results are believed to be very accurate: the mapping described above seems to break down somewhere below the mean field threshold.
As our data are not accurate in this limit, we are not able to detect the point where this breaking occurs.
Moreover the relation noted here, given in eq. (\ref{eq:mappingrrf}), corresponds to the famous one $D_{RF} \rightarrow D_{RF}-2$ that is found for short range $D$-dimensional models if we use the usual mapping between these models and long range one dimensional models  \cite{LarKatzMoo}, \cite{BanosFernMartYou}, that is thought to hold in the mean field region and in the vicinity of the mean field threshold. 
Thus, the apparent breakdown of this relation is similar to what happens in $D$-dimensional short range models \cite{Aharony, YoungPer}, \cite{ParSour}, as we described in the introduction \cite{TarjTiss1, TarjTiss2,TarjTiss3,TarjusTissierBalog}.

\section{Summary and Conclusions}

In this work we numerically computed the values of $1/\nu$ for different values of $\rho$ in the RFHM, exploring the whole non-mean field region.
We have then found an analytic expression for the critical exponent $\delta(\rho)$, see eq. (\ref{eq:tre_delta}). 
This expression, with the correspond one computed for the pure HM in \cite{KimThom}, eq. (\ref{eq:fit_delta_pure}), leads to eq. (\ref{eq:mapping}) and eq. (\ref{eq:mappingrrf}) between the critical exponents of the RFHM and of the pure HM that appears to hold exactly for the $\delta$ but not for the other exponents.

Let's also compare our result with the $D$-dimensional RFIM. As was accurately studied in \cite{BanosFernMartYou} and discussed in \cite{ParisiLeuzzi} it's possible to compare these two models not too far from the upper critical dimension, that is $6$ for the RFIM, and thus we don't expect to have a satisfying agreement between the RFIM results in $D=3$ and the corresponding ones in the RFHM. In fact, we don't. 
In \cite{MayorFytas} Martìn-Mayor and Fytas computed $\overline{\eta}=1.0268(1)$ and $\nu=1.34(11)$ for the RFIM in $D=3$. The $\rho$ corresponding to $D=3$ is $\rho=(2-\overline{\eta}_{SR}/2)/D+1= 1.49550(2)$, \cite{ParisiLeuzzi}. What we should compare is $\nu_{LR}(\rho=1.49550(2))=14(3)$ and $D \nu_{SR} (D) = 4.0(3)$ for $D=3$ and it's clear that they don't agree.
The situation improves in $D=4$, where from the results of \cite{Middleton} we get $\nu=0.82(6)$ and $\overline{\eta}=0.45(17)$. The $\rho$ corresponding to $D=4$ is $\rho=1.444(21)$ at which we get $\nu_{LR}(\rho=1.444(21)) = 4.2(8)$ and this result has to be compared to 
$D \nu_{SR} (D) =3.3(2) $ for $D=4$.

Another interesting comparison can be done with the critical exponents of the long range model evaluated in \cite{ParisiLeuzzi} and \cite{Hartmann}. In $\rho=1.25$ our result $1/\nu=0.260(6)$ has to be compared with the long range value $1/\nu = 0.262 ± 0.035$ obtained in \cite{Hartmann}. Thus they are found to be in good agreement as expected because $\rho=1.25$ belongs to the mean field region, where we expect $1/\nu=0.25$.
In $\rho=1.4$ we found $1/\nu=0.288(4)$ and it has to be compared with $1/\nu = 0.316(9)$\cite{ParisiLeuzzi} and $1/\nu = 0.29(3)$\cite{Hartmann}. and even if it's not clear if we should find the same value, they seems to be quite similar.

\section{Acknowledgements}

The research leading to these results has received funding from the European Research Council under the European Union's Seventh Framework Programme (FP7/2007-2013) / ERC grant agreement n° [247328].
We would also like to acknowledge many stimulating discussions with F. Metz and A. Decelle.

\end{document}